\documentclass[onecolumn,notitlepage,nofootinbib,floatfix,amsmath,showpacs,showkeys,aps,pre]{revtex4-1}
\usepackage[utf8]{inputenc}
\usepackage[final]{graphicx}
\usepackage{bm}
\usepackage[normalem]{ulem}
\usepackage[usenames]{color}
\usepackage{times}
\usepackage{amsmath}
\usepackage{verbatim}
\usepackage{amssymb}
\usepackage{xcolor}
\usepackage{hyperref}
\usepackage{physics}
\usepackage{wrapfig}
\usepackage{subfigure}
\DeclareGraphicsExtensions{.eps}

\begin{document}

\title{Effects of the kinetic energy in heat for overdamped systems}


\author{Pedro V. Paraguass\'{u}}
 \email{paraguassu@aluno.puc-rio.br}
\affiliation{Departamento de F\'{i}sica, Pontif\'{i}cia Universidade Cat\'{o}lica\\ 22452-970, Rio de Janeiro, Brazil}

\author{Rui Aquino}
\affiliation{Departamento de F\'{i}sica Te\'{o}rica\\ Universidade do Estado do Rio de Janeiro\\ 20550-013, Rio de Janeiro, Brazil}

 \author{Lucianno Defaveri}
 \affiliation{Department of Physics, Bar Ilan University, Ramat-Gan 52900, Israel}

\author{Welles A.~M. Morgado}
 \email{welles@puc-rio.br}
\affiliation{Departamento de F\'{i}sica, Pontif\'{i}cia Universidade Cat\'{o}lica\\ 22452-970, Rio de Janeiro, Brazil\\ and National  Institute of Science and Technology for Complex Systems}

\date{April 2022}

\begin{abstract}
In the derivation of the thermodynamics of overdamped systems, one ignores the kinetic energy contribution, since the velocity is a slow variable. In this paper, we show that the kinetic energy needs to be present in the calculation of the heat distribution to have a correct correspondence between the underdamped and overdamped cases, meaning that the velocity can not be fully ignored in the thermodynamics of these systems. We do this by investigating in detail the effect of the kinetic energy for three different systems, the harmonic potential, the logarithm potential, and an arbitrary non-isothermal process.

\end{abstract}

\maketitle

\section{Introduction}

The attempt to understand and control heat has been pursued by humanity since before the dawn of modern civilization \cite{barnett1946development,barnett1946development2}. Nowadays, we currently understand heat as a disordered form of energy that can be used as a source for thermal machines \cite{holubec2021fluctuations,ciliberto2017experiments}. With the development of new technologies, we can now manipulate small systems, on the order of nanometers, to the point of building such thermal machines on this scale \cite{martinez2016brownian,blickle2012realization}.

On the microscale,  thermodynamics has a different character than in the macro-world. Namely, quantities like heat and work now become fluctuating quantities. A lot of attention was devoted to obtaining the probability distribution for these thermodynamic functionals, with experimental \cite{joubaud_fluctuation_2007,gomez-solano_heat_2011,imparato_work_2007,imparato_probability_2008}  and theoretical \cite{chatterjee_exact_2010,chatterjee_single-molecule_2011,ghosal_distribution_2016,goswami_heat_2019,paraguassu_heat_2021,chvosta_statistics_2020,chen2021quantum,paraguassu2022heat} calculations.
This microscale thermodynamics is now a proper field of research, that is now called stochastic thermodynamics \cite{peliti2021stochastic,sekimoto2010stochastic,seifert2012stochastic}.

 Most of the works in stochastic thermodynamics make use of overdamped models, where one can neglect the inertia of the particle due to the large dissipation from contact with a heat bath. When this is the case, one in principle neglects the kinetic energy, by disregarding the inertia, as opposed to the underdamped case, where we always have this contribution \cite{sekimoto2010stochastic}. However, since there is a distinction between the probability distributions when $m=0$ is taken first, instead of taking $m/\gamma\rightarrow0$ after the calculations~\cite{nascimento2019}, the kinetic energy needs to be present. Moreover, the velocities in a overdamped system are stationary, obeying a Boltzmann Gibbs distribution.

Futhermore, in \cite{arold2018heat} it has been shown that for overdamped systems in an isochoric process, the kinetic energy can not be ignored. They have shown that the average heat of the overdamped Brownian motion, when there is a temperature protocol, does not correspond to the underdamped case in the large friction limit. At the same time, if no temperature protocol is present they find that the correspondence between the averages is satisfied. The problem is to take the limit in the system dynamics before the calculation of its thermodynamics. While we can ignore the velocity in the dynamics and in the average thermodynamics for isothermal processes, in the fluctuating thermodynamics the kinetic energy needs to be present. A conclusion already found by Ref. \onlinecite{garcia2018heat}.

Another result that breaks the correspondence between underdamped and overdamped systems was obtained by the authors in \cite{paraguassu_heat_2021_2}. We have showed that the heat distribution of the free particle in the underdamped case, \begin{equation}
    P_u(Q) =\frac{\sqrt{\coth \left(\frac{\gamma  \tau }{m}\right)+1}}{\pi \sqrt{2}T} \;K_0\left(\frac{| Q|  \sqrt{\coth \left(\frac{\gamma  \tau }{m}\right)+1}}{\sqrt{2} T}\right),\label{under}
\end{equation}
does not correspond to the heat distribution of the free particle in the overdamped case 
\begin{equation}
    P_o(Q) = \delta (Q),\label{free}
\end{equation}
in the limit of large friction, where $\gamma >> m$. The absence of the correspondence manifest the necessity of considering the kinetic energy. 

In this paper, we show that by considering the kinetic energy, the correspondence is satisfied between the heat distributions. Different from the conclusions in \cite{arold2018heat} our result shows that even in an isothermal process, there is the necessity to consider the kinetic energy in order to obtain the correct heat distribution of an overdamped system.

The aim of this paper is to show that the presence of the kinetic energy is necessary concerning the thermodynamics of overdamped systems, due to the necessity of a correspondence between the heat distribution of an underdamped system and the corresponding overdamped one. We start by calculating explicitly the correct heat distribution for the free particle case. Then, we investigate the differences when the kinetic energy is included for harmonic and logarithm systems. We find that there are in general more fluctuations. Moreover, since the kinetic energy is more relevant for non-isothermal processes, we also investigate the heat of a free particle subject to an arbitrary protocol, where the only constraint is that the initial and final temperatures are well-defined.

In section II, we review the free-particle case, showing the fundamental differences when including the kinetic energy. In section III and IV, the heat distribution for the harmonic and logarithmic  cases are obtained respectively, and studied in details its fluctuations by the calculation of the central moments. In section V, we study the effect of considering the kinetic energy in a non-isothermal process. In section VI, we present our conclusions and discussions.

\section{The kinetic energy}

A free Brownian particle in contact with a heat bath in the overdamped regime obeys the Langevin equation (always assuming $k_B=1$)
\begin{equation}
    \gamma \dot x(t) = \eta(t),
\end{equation}
where $\gamma$ is the friction coefficient, and $\eta(t)$ is a white noise process, with zero mean and correlation $\langle\eta(t)\eta(t') \rangle = 2\gamma T \delta(t-t')$, where $T$ is the temperature of the heat bath. 
 
If one considers the definition of heat given by Sekimoto \cite{sekimoto2010stochastic}, 
\begin{equation}
    Q[x]=\int_0^\tau \left(\gamma\dot x - \eta(t)\right) \dot x dt,\label{seki}
\end{equation}
which is the definition of the energy exchanged between the Brownian particle and the heat bath, one obtains that $Q[x]=0$ in the free case. One can ask if this result is compatible with our physical intuition. To move, the particle has to receive or lose energy in some way. What $Q[x]=0$ is saying is that there are no fluctuations in the energy exchanged between the particle and the heat bath. Using the definition in Eq.~\ref{seki}, the heat distribution of the free particle in the overdamped case has a simple formula given by a Dirac delta, as shown in Eq.~\ref{free}, which is a deterministic probability (in a sense that there is no fluctuation). In addition, if one calculates the heat distribution of the free particle in the underdamped regime, one obtains Eq.~\ref{under} \cite{paraguassu_heat_2021_2}. The problem is, Eq.~\ref{under} does not correspond to Eq.~\ref{free} in the large friction limit $\gamma >> m$. If we take this limit, we obtain
\begin{equation}
    P_u(Q) \xrightarrow[\gamma >> m]{} \frac{1}{\pi T}\; K_0\left(\frac{|Q|}{T}\right).
\end{equation}
What we find is that the solution of this problem is solved by considering the kinetic energy in the heat of the overdamped system. For the free particle case, this means that the heat is now
\begin{equation}
    Q[x] = \frac{1}{2}m \left(v_\tau^2-v_0^2\right)=\Delta K,
\end{equation}
where $v_\tau$ is the final velocity and $v_0$ the initial velocity. By calculating the distribution of this heat we can recover the correspondence. The calculation is easy since in the overdamped limit we assume that the velocities obey a stationary equilibrium distribution at all times. This means that the initial distribution and transitional distribution of the particle velocities are
\begin{equation}
    P(v_0)P[v_\tau,\tau|v_0,0] = \frac{m }{2\pi T}e^{-\frac{m}{2T}(v_\tau^2+v_0^2)}.
\end{equation}
Thus, if we calculate the heat distribution, we will have the characteristic function (see the appendix)
\begin{eqnarray}
    Z(\lambda) &=& \int dv_0 \int dv_\tau \frac{m }{2\pi T}e^{-\frac{m}{2T}(v_\tau^2+v_0^2)}e^{-i\lambda\frac{m}{2}(v_\tau^2-v_0^2)}\nonumber\\
    &=& \frac{1 }{\sqrt{1+\lambda ^2T^2}},\label{charc}
\end{eqnarray}
that can be Fourier transformed to obtain the desired heat distribution
\begin{equation}
    P_o (Q)= \int \frac{d\lambda}{2\pi}\frac{e^{i\lambda Q}}{\sqrt{1+\lambda ^2T^2}}=\frac{1}{\pi T}\; K_0\left(\frac{|Q|}{T}\right),\label{correct}
\end{equation}
which is the exact heat distribution that one obtains if one calculates the underdamped heat distribution and takes the large friction limit. Also, note that the heat distribution is stationary, due its independence in time.

As a result, to have a correspondence between the heat distributions of the two cases, one needs to take into account the kinetic energy term \cite{garcia2018heat}. Moreover, this means that the definition of heat given by Sekimoto Eq.~\ref{seki} is probably not the complete version of the heat in stochastic thermodynamics. Nevertheless, a complete version could be achieved if one can always define the heat accordingly to the first law using the complete energy of the system, i.e., the potential energy plus the kinetic energy. The natural recipe is: firstly defining the work, and then using the first law together with the complete energy of the system, define the heat. Although the distributions in equations \ref{free} and \ref{correct} have a similar statistical behavior, both distribution have a singularity in $Q=0$ and gives the same average for the heat $\langle Q[x]\rangle=0$. Furthermore, in the overdamped limit in an isothermal process, where the velocities are in equilibrium, the average kinetic energy is always zero. Explaining why in \cite{arold2018heat} the average heat in an isothermal process has the correspondence between its overdamped and underdamped cases.

\section{Harmonic system}

\begin{figure}
    \centering
    \includegraphics[width=8.6cm]{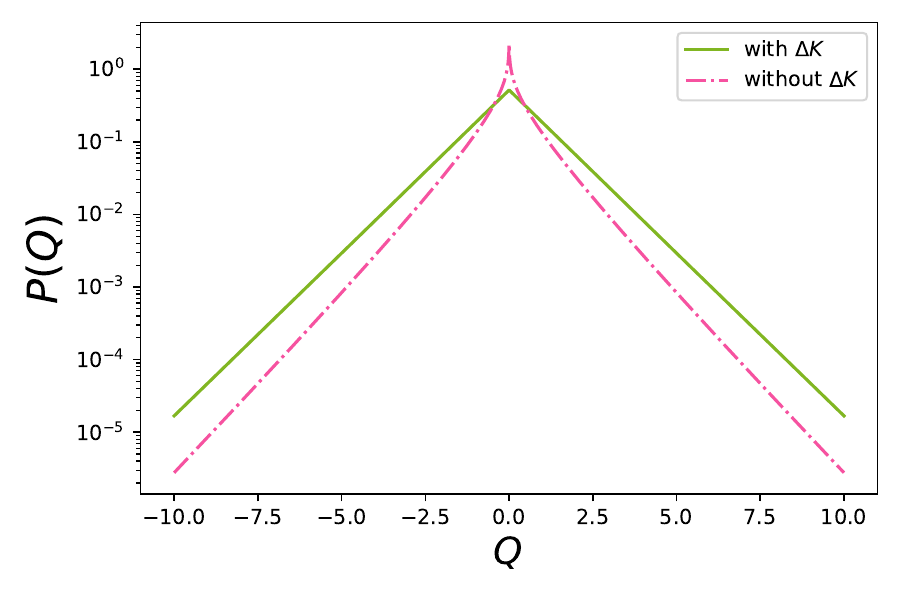}
    \caption{Heat distribution with and without the kinetic energy. The pink dashed line it is the case without the kinetic energy, while the solid green line is the corrected heat distribution with the kinetic energy. All the constants are set to one. We can see that the correct heat distribution allows more fluctuations for the heat due the velocities.}
    \label{fig}
\end{figure}

The necessity to include the kinetic energy, modifies some results already obtained in the literature. Here, we investigate the effects of the kinetic term in the harmonic potential case that was first calculated (without the kinetic contribution) in \cite{chatterjee_exact_2010}. This system obeys the Langevin equation
\begin{equation}
    \gamma \dot x(t) = k x(t) + \eta(t),
\end{equation}
where $k$ is the stiffness of the harmonic potential. We start by noticing that by considering the kinetic energy we only add more integrals in the calculation of the characteristic function. That is
\begin{equation}
    Z(\lambda) = \int dv_0 \int dv_\tau \frac{m }{2\pi T}e^{-\frac{m}{2T}(v_\tau^2+v_0^2)}e^{-i\lambda\frac{m}{2}(v_\tau^2-v_0^2)} z(\lambda),\label{eq11}
\end{equation}
where $z(\lambda)$ is the characteristic function that one obtains if we do not consider the kinetic energy. For the harmonic case, this characteristic function is \cite{chatterjee_exact_2010,paraguassu_heat_2021_2}
\begin{equation}
    z(\lambda) = \frac{\sqrt{\left(\coth \left(\frac{k \tau }{\gamma}\right)+1\right)}}{\sqrt{\coth \left(\frac{k  \tau }{\gamma}\right)+2 \lambda ^2 T^2+1}}.
\end{equation}
Integrating in the velocities, we obtain the correct characteristic function
\begin{equation}
    Z(\lambda) = \frac{1 }{\sqrt{1+\lambda ^2T^2}}\frac{\sqrt{\left(\coth \left(\frac{k \tau }{\gamma}\right)+1\right)}}{\sqrt{\coth \left(\frac{k  \tau }{\gamma}\right)+2 \lambda ^2 T^2+1}},
\end{equation}
that, unfortunately, does not allow us to obtain analytically the heat distribution $P(Q)$. Nevertheless, we can integrate it numerically, and the result is showed in figure~\ref{fig}, in comparison with the heat distribution obtained without the kinetic energy correction. What we find is that the corrected heat distribution has larger fluctuation tails than the naive distribution. This occurs due to the equilibrium fluctuations of the velocities. Having a greater probability of occurring values of heat far from the mean, it can be exploited in the design of thermal machines \cite{holubec2021fluctuations}, where one wants to use the fluctuations to improve the efficiency of these machines.

In the asymptotic time limit of the system, we can recover an analytical solution by noticing that 
\begin{equation}
    \lim_{\tau\rightarrow \infty} Z(\lambda) =  \frac{1}{(\lambda^2T^2+1)},
\end{equation}
which gives an exponential distribution for the heat distribution
\begin{equation}
     \lim_{\tau\rightarrow \infty} P_o(Q) = \frac{e^{- \frac{| Q|}{T} }}{2T}.
\end{equation}
A result different from the Bessel function obtained in  \cite{imparato_work_2007}, $P_{\text{Imparato}}(Q) \sim K_0( T^{-1}|Q|) $. Moreover, this asymptotic heat distribution is the same asymptotic distribution of the harmonic case in the underdamped regime \cite{paraguassu_heat_2021_2}. Therefore, in the asymptotic limit, the heat distribution is the same for underdamped and overdamped systems in a harmonic potential.

\subsection{General considerations}

Some general considerations can be obtained by the characteristic function formula. It is easy to see that for an isothermal overdamped process the new characteristic function will always be
\begin{equation}
    Z(\lambda) = \frac{z(\lambda)}{\sqrt{1+\lambda^2T^2}},
\end{equation}
where this is valid for system with velocity independent of the position. Due the quadratic dependence in $\lambda$ in the new term, the mean and skewness are no affect by the kinetic energy. That is, $\langle Q \rangle = \langle Q \rangle^{\text{kin}}$. Moreover the difference between the variances will always be given by
\begin{equation}
    \sigma_Q^{\text{kin}}-\sigma_Q = T^2\label{vargen}
\end{equation}
where the superscript $\rm kin$ is used to specify that we are obtained the variance with the new characteristic function, and the moments can always be calculated by 
\begin{equation}
    (-i)^n  \frac{\partial^n Z(\lambda)}{\partial \lambda^             n}\bigg|_{\lambda=0}=\langle Q^n \rangle.
\end{equation}
This difference means that, for higher temperatures, the difference between the variances increases. Thus the effect of including the kinetic energy is more visible for high temperatures.
As far as we known, for the kurtosis no general statement appears.

\subsection{Central Moments}

Now, let us study in detail the fluctuations of the heat with harmonic potential. We will compare the central moments; mean, variance, skewness, and kurtosis, with and without the kinetic energy. 

The odd moments, the mean and skewness are null for both cases, since the distribution is symmetric around zero and the kinetic energy does not affect this behavior. 

For the variance, we have a slight difference, while the variance without kinetic energy is
\begin{equation}
    \sigma_Q = \frac{2 T^2}{\coth \left(\frac{k \tau }{\gamma }\right)+1},
\end{equation}
the case with kinetic energy is
\begin{equation}
    \sigma_Q^{\text{kin}} = T^2 \left(2-e^{-\frac{2 k \tau }{\gamma }}\right).
\end{equation}
Where one can check that these variances satisfy Eq.~\ref{vargen}.

For the kurtosis, the fourth central moment, we have
\begin{equation}
\kappa = \frac{\langle(Q-\langle Q \rangle)^4 \rangle}{\sigma_Q^2}= 9, \;\;\; \kappa^{\rm kin} = \frac{3}{\left(1-2 e^{\frac{2 k t}{\gamma }}\right)^2}+6,
\end{equation}
where the constant value for the kurtosis without kinetic energy comes from the variance having the same dependency of the constants as the fourth moment. Both cases are always Leptokurtic \cite{westfall2014kurtosis}, since the kurtosis is greater than 3 for all times. 

\section{Logarithm system}

Here we show the modification of another interesting case of the heat distribution in the logarithm potential calculated without correction by one of the authors in \cite{paraguassu_heat_2021}. The logarithm potential appears in different stochastic phenomena \cite{leibovich_aging_2016,hirschberg_approach_2011,aghion_non-normalizable_2019,dechant_superaging_2012,barkai_area_2014,kessler_infinite_2010,ray_diffusion_2020}, and its Langevin equation is (we take $\gamma=1$ for simplicity)
\begin{equation}
     \dot x(t) = \frac{k}{x(t)} + \eta(t),
\end{equation}
where now $k$ is the strength of the logarithm potential and now $x(t)$ is defined only in the positive real axis \cite{ryabov2015stochastic}, and now the initial distribution of the position is a Dirac delta \cite{paraguassu_heat_2021}. Now we will explicitly show the modifications that one have to consider when calculate the heat distribution. This time,
without using the characteristic function method, and choosing to illustrate another approach, where instead of the Fourier transform of the Dirac delta, we use its properties.

\begin{figure}
    \centering
    \includegraphics[width=8.6cm]{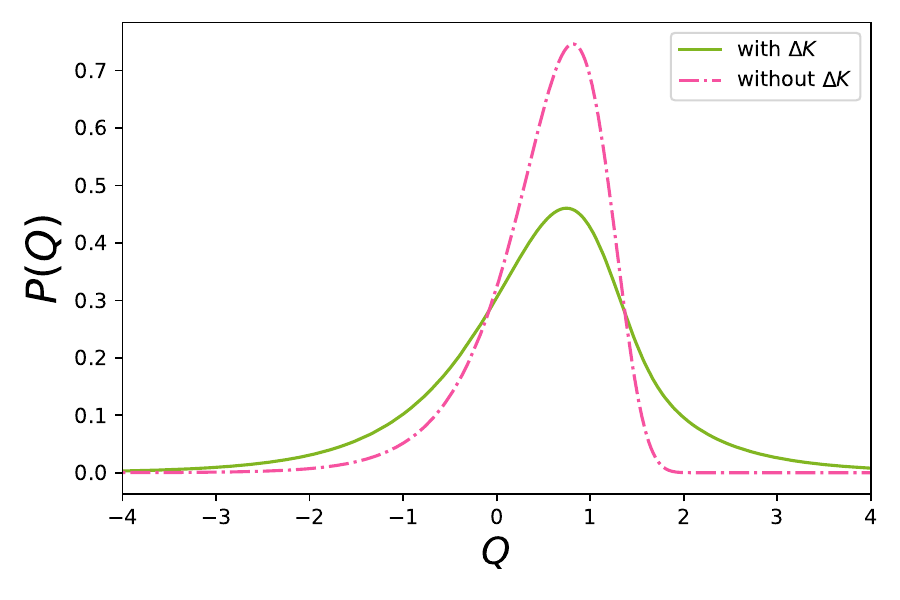}
    \caption{Probability distribution of the heat for the logarithm case, with and without the kinetic energy. The pink dashed line it is the case without the kinetic energy, while the solid green line is the corrected heat distribution with the kinetic energy. All constants are set to one. Note that the kinetic energy has a more broaden distribution, as expected since the variance increases due the kinetic energy.}
    \label{fig2}
\end{figure}

The heat is shifted with the kinetic correction, therefore we have
\begin{equation}
    Q[x] = \Delta K + k \ln\frac{x_\tau}{x_0},
\end{equation}
and thus the heat distribution is
\begin{equation}
    P(Q) = \Bigg\langle \delta\left(Q - \Delta K - k \ln\frac{x_\tau}{x_0}\right)\Bigg\rangle.
\end{equation}
If we define $\Tilde{Q}=Q-\Delta K$, the calculation can be carried along the same lines  in reference \cite{paraguassu_heat_2021}. However we will still need to integrate the velocities. Hence, we arrive at the integrals
\begin{equation}
    P(Q) = \int dv_0 dv_\tau\frac{m }{2\pi T}e^{-\frac{m}{2T}(v_\tau^2+v_0^2)}p\left(Q-\Delta K\right),\label{eq24}
\end{equation}
where $p(Q)$ is the expression of the heat distribution obtained in \cite{paraguassu_heat_2021} without the kinetic energy. Note that, the above expression is similar to one found by \cite{garcia2018heat}. Again we cannot obtain an analytical result for the heat, but it can be solved by numerical integration. The result is plotted in figure~\ref{fig2}, where one can see that the distribution with kinetic energy has a more broaden probability. Furthermore, it is interesting to note that previous analytical results, when considering the correction of the kinetic energy, not give anymore an analytical solution. This suggests that the heat distribution of overdamped systems is way more complex when the kinetic energy is taken into account. 

\subsection{Central Moments}

Different from the harmonic case, the logarithm potential does not have an equilibrium initial distribution or null odd moments. Here we compare the central moments of heat between the case with and without kinetic energy. 

The characteristic function of the logarithm case without the kinetic energy is \cite{paraguassu_heat_2021,paraguassu2022heat},
\begin{eqnarray}
   z(\lambda)= 2^{i k \lambda } x_0^{-i k \lambda } (t T)^{\frac{i k \lambda }{2}} \Gamma \left(\frac{i T \lambda  k+k+T}{2 T}\right) \, _1\tilde{F}_1\left(-\frac{1}{2} i k \lambda ;\frac{k+T}{2 T};-\frac{x_0^2}{4 t T}\right),
\end{eqnarray}
where, $_1\tilde{F}_1$ is the  hypergeometric regularized function \cite{abramowitz1988handbook}, while the characteristic function with the kinetic energy is just $Z(\lambda) = z(\lambda)(1+\lambda^2T^2)^{-1/2}$. Therefore, the moments can be calculated analytically.

\begin{figure}
    \centering
    \includegraphics[width=8.6cm]{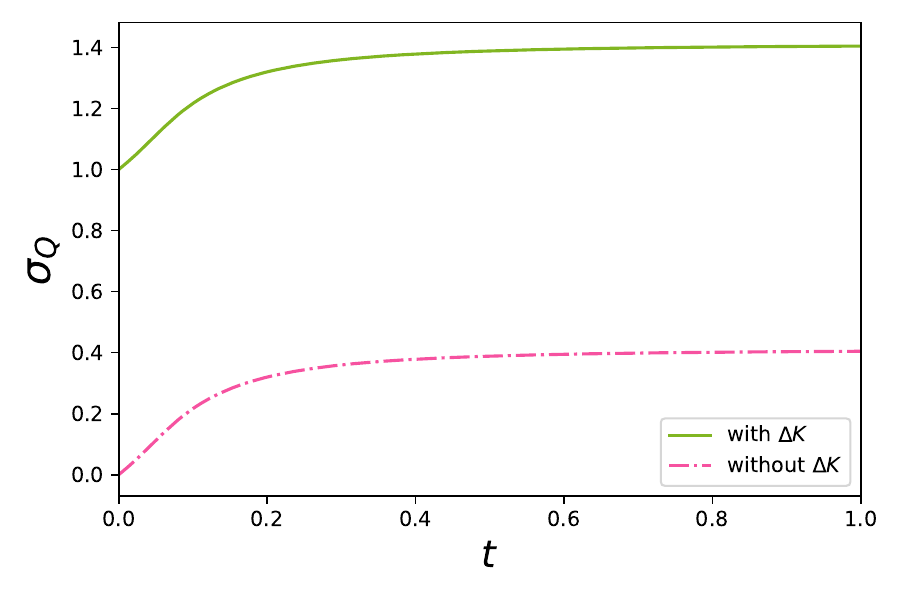}
    \caption{Variance with and without kinetic energy for the logarithm potential case. All constants are set to one.}
    \label{variance}
\end{figure}

The odd central moments; the mean and skewness are exactly the same, since the new term due the kinetic is a quadratic dependence in $\lambda$. The mean is given by
\begin{eqnarray}
     \langle Q\rangle =\langle Q\rangle^{\text{kin}}= \frac{1}{2} k \left(-\Gamma \left(\frac{k+T}{2 T}\right) \; _1\tilde{F}_1^{(1,0,0)}\left(0,\frac{k+T}{2 T},-\frac{x_0^2}{4 t T}\right)+\psi ^{(0)}\left(\frac{k+T}{2 T}\right)+\log \left(\frac{4 t T}{x_0^2}\right)\right),
\end{eqnarray}
where $\psi$ is the Polygamma function. The mean is increasing as time pass, as already showed in \cite{paraguassu_heat_2021}. The variance is given by
\begin{eqnarray}
    \sigma_Q^{\rm kin} = \sigma_Q + T^2 = \frac{1}{4}\left(4 T^2+k^2 \left(\log (t T) \log \left(\frac{16 t T}{x_0^4}\right)-\left(\log \left(\frac{16 t T}{x_0^2}\right)-2 \log (x_0)\right) \left(\log \left(\frac{t T}{x_0^2}\right)+2 \log (x_0)\right)\right)-\right.\nonumber\\+k^2 \psi ^{(1)}\left(\frac{k+T}{2 T}\right)- \left.k^2 \Gamma \left(\frac{k+T}{2 T}\right)^2 _1\tilde{F}_1^{(1,0,0)}\left(0,\frac{k+T}{2 T},-\frac{x_0^2}{4 t T}\right)^2+k^2 \Gamma \left(\frac{k+T}{2 T}\right) _1\tilde{F}_1^{(2,0,0)}\left(0,\frac{k+T}{2 T},-\frac{x_0^2}{4 t T}\right)\right).
\end{eqnarray}
Despite this complicated formula the variance has a simply behavior as showed in figure~\ref{variance}. The variance also increase as time pass until reach a stationary behavior. 

\begin{figure}
    \centering
    \includegraphics[width=8.6cm]{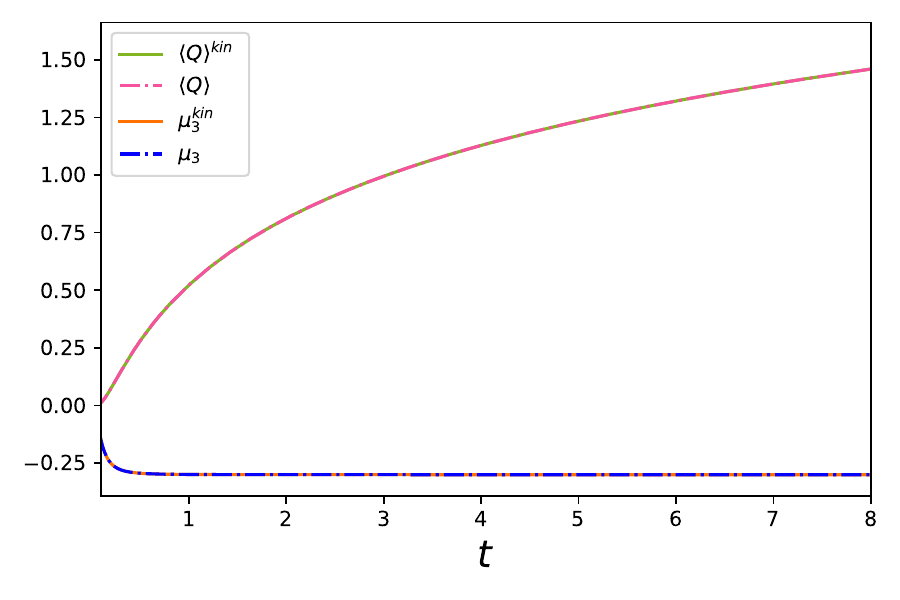}
    \caption{Mean and skewness for the logarithm case with and without the kinetic energy. All parameters are set to one. Note that the mean is always increasing, while the skewness becomes constat. }
    \label{fig3}
\end{figure}

\begin{figure}
    \centering
    \includegraphics[width=8.6cm]{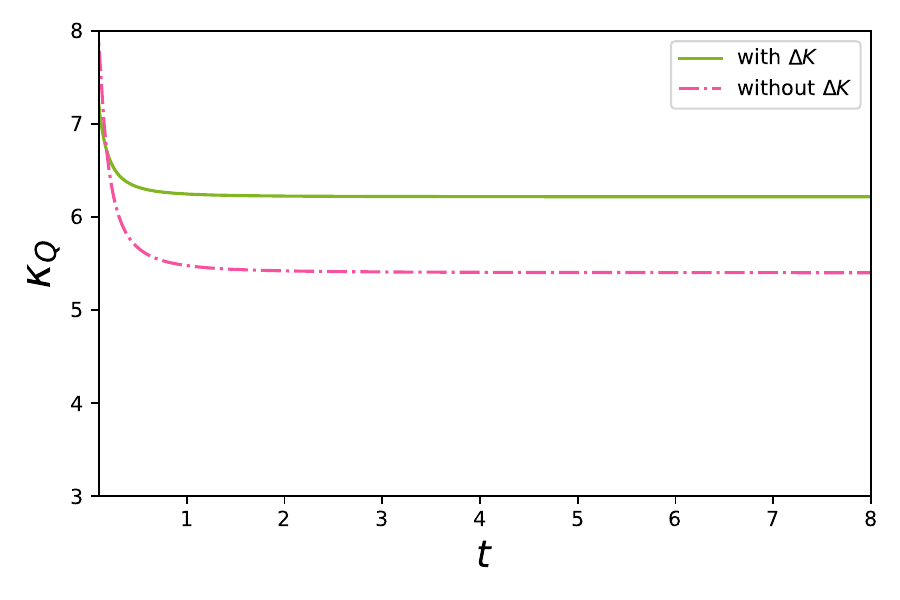}
    \caption{kurtosis for the logarithm case, with and without kinetic energy. All constants are set to one.}
    \label{fig4}
\end{figure}

The next central moments are very long expressions, and the results are plotted in figures \ref{fig3} and \ref{fig4}. For the skewness one can see that it is negative, meaning that the left tail of the distribution is longer than the right tail, measuring the asymmetry of the distribution. The dynamics is simply, the skewness rapidly saturates in its stationary value.

The behavior of the kurtosis is more complex. For the case without kinetic energy the kurtosis saturates very rapidly in a fixed value greater than 3, meaning that the distribution without kinetic energy is leptokurtic, while for the case without kinetic energy, the tail is also became leptokurtic, however with a small value. Therefore, as expected, the tails with kinetic energy are more fatter.

\section{Non-isothermal Process}

From the theoretical point of view, we see that taking the kinetic energy into account preserves the consistency of stochastic thermodynamics. From a practical point of view, however, the fluctuations in kinetic energy will not be dominated by the protocol, but by the accuracy of the apparatus. 
The case with isothermal is simple and does not have substantial consequences in experimental measures. However, for the non-isothermal process, the effect of including the kinetic energy is more apparent, as already shown in \cite{arold2018heat}. As a first model, here we will use the free particle case connected with two different equilibrium heat baths in the initial and final time of the process. 

The position of an overdamped free particle in contact with a heat bath obeys again the following Langevin equation 
\begin{equation}
    \gamma \dot x(t) = \eta(t),
\end{equation}
where now $T(t)$ is the time-dependent temperature of the heat bath. Here, we are considering a temperature protocol
\begin{equation}
    T(t) = T_0 + (T_f-T_0)f(t), \label{protocol}
\end{equation}
where we are assuming $f(t)\rightarrow 1$ in the end of the process in order to give $\langle v_\tau^2\rangle \sim T_f$. That is, the final velocity is in equilibrium with the final state of the heat bath with temperature $T_f$. Where or not this equilibration is possible escape from the scope of this paper and will be addressed in a future work.

The heat exchanged by the particle will be just the kinetic energy
\begin{equation}
   Q =  \Delta K = \frac{1}{2}m (v_\tau^2-v_0^2),
\end{equation}
Note that, if the kinetic energy os not considered there is no meaning in discussing the heat for this system.

The characteristic function of the heat can be calculated along the same lines of the previous cases, assuming that the particle velocity equilibrates at the end of the process, we have
\begin{eqnarray}
    Z(\lambda) &=& \int dv_0 \int dv_\tau \frac{m }{2\pi \sqrt{T_0T_f}}e^{-\frac{mv_\tau^2}{2T_f}-\frac{mv_0^2}{2T_0}}e^{-i\lambda\frac{m}{2}(v_\tau^2-v_0^2)}\nonumber\\
   &=& \frac{1}{\sqrt{1+i\lambda T_f}}\frac{1}{\sqrt{1-i\lambda T_0}}.\label{equili}
\end{eqnarray}
Note that, the Fourier transform of the characteristic function gives the Bessel function in the probability distribution for the heat, when $T_0=T_f$ as showed in \cite{chatterjee2010exact}.
Furthermore the probability distribution can be calculated analytically (see appendix \ref{apA}), and the result is
\begin{equation}
    P(Q) = \frac{1}{\pi}e^{\frac{\Delta T}{2T_0T_f}} K_0\left(|Q|\sqrt{\frac{\Delta T^2 +4T_0T_f}{2T_0T_f}}\right).\label{distheat}
\end{equation}
It should be noted that we have an exponential factor outside the Bessel function, that depends on the two temperature difference.

Surprisingly, the characteristic function obeys a kind of exchange fluctuation theorem for the heat \cite{jarzynski2004classical}.
\begin{equation}
    Z\left(i\left(\frac{1}{T_f}-\frac{1}{T_0}\right)\right) = 1,\label{EFT}
\end{equation}
It is surprising because in \cite{jarzynski2004classical} the setup studied is different from our system. While in \cite{jarzynski2004classical} the setup is two bodies in contact with two different reservoirs, here we have one body (the free particle) in contact with a heat bath that changes its temperature.

Moreover, as a matter of curiosity, we have an interesting result if we choose the initial temperature in the characteristic function
\begin{equation}
    Z\left(-iT_0^{-1}\right) = \frac{1}{\sqrt{2\eta_C}},
\end{equation}
where $\eta_C$ is the Carnot efficiency. As far as we known, there is no application of this formula. Nevertheless we can see this as a bound in the fluctuations of the heat. Since $Z\left(-iT_0^{-1}\right)=\langle e^{-Q/T_0} \rangle$, we have
\begin{equation}
    \langle e^{-Q/T_0} \rangle = \frac{1}{\sqrt{2\eta_C}}\leq e^{-\langle Q\rangle/T_0},
\end{equation}
which follows from the Jenssen inequality.

\subsection{Heat Statistics}

The first moment is the mean, and is given by
\begin{equation}
    \langle Q \rangle = \frac{T_f-T_0}{2}.
\end{equation}
This result is thermodynamically consistent because when we assume $T_f>T_0$, that is, the temperature of the bath is increased after the protocol, we have $ \langle Q \rangle >0$, meaning that the particles gain energy if we increase the temperature. Therefore, the heat flows from a hotter (the bath) to a colder body (the particle).

The variance will be
\begin{equation}
    \sigma^2_Q =  \frac{1}{2}\left(T_f^2+T_0^2\right).
\end{equation}
Being a measure of dispersion, the variance of the heat is greater for greater temperatures and does not is affected by the signal difference between $T_0$ and $T_f$.

The next quantity is the skewness which measure the asymmetry of the distribution. By simplicity we use the non-normalized skewness, which is just calculated by $ \mu_3 = {\langle \left(Q-\langle Q \rangle\right)^3\rangle}. $
Thus, using the characteristic function we have
\begin{equation}
    \mu_3 = 
    {T_f^3 - T_0^3}\label{skefree}
\end{equation}
We obtain a negative skewness for $T_0>T_f$, meaning that negative values of heat are more probable, which is what we expected due to thermodynamic consistency. For $T_0<T_f$ the behavior is the opposite. 

The last quantity is the  kurtosis, which informs about the tail extremity of the distribution \cite{westfall2014kurtosis}. We define the excess kurtosis as
\begin{eqnarray}
    \kappa_Q &=& \frac{\langle \left(Q-\langle Q \rangle\right)^4\rangle}{\left(\sigma_Q^2\right)^2}\\&=& \frac{3 \left(5 T_0^4+2 T_0^2 T_f^2+5 T_f^4\right)}{\left(T_0^2+T_f^2\right)^2},\label{kurtfree}
\end{eqnarray}
which is always greater than 3 for positive values of $T_0$ and $T_f$, meaning that the distribution is Leptokurtic, i.e. the distribution has fatter tails compared with a normal distribution, similar with the heat distribution for a particle in a logarithm-harmonic potential \cite{paraguassu2022heat}.  Having fatter tails means that we have more chance to occur events far from the mean

For a non-isothermal process, we thus find that, by including the kinetic energy, the found heat distribution leads to consistent results with thermodynamics.

\section{Conclusion}

By considering the kinetic energy, some theoretical works might need to be revisited. Mainly, the works on the fluctuations of the heat in the isothermal process \cite{chatterjee_exact_2010,chatterjee_single-molecule_2011,ghosal_distribution_2016,goswami_heat_2019,paraguassu_heat_2021,chvosta_statistics_2020} might need to be adjusted to include the kinetic energy. That is, those calculations need to be taken using Eq.~\ref{eq11} and \ref{eq24}. Furthermore, some results in harmonic thermal machines need to be reviewed to properly calculate the fluctuations of the efficiency of such machines. We have shown that the heat distribution with the kinetic energy allows more fluctuations for the exchanged heat, a result that can be exploited in the development of thermal machines. Moreover, it ensures that by considering the kinetic energy, the calculation of the heat distribution becomes more difficult. Previous analytical results, obtained without the kinetic energy, are not possible now, as an example here we have shown the harmonic and logarithm potentials. 

It will be interesting to see whether there are  cases where we can take the overdamped limit before the calculation of the thermodynamic properties. One case was already shown in \cite{arold2018heat}, where the average heat is not affected by taking the overdamped limit before in an isothermal process.

In conclusion, the inclusion of kinetic energy allows us to keep the theoretical consistency of stochastic thermodynamics since it validates the correspondence between the underdamped and overdamped heat distribution. Here we investigate in detail the effects of the kinetic energy for harmonic, logarithm, and an arbitrary non-isothermal process. We calculated the central moments and the heat distribution, seeing that the fluctuations of the heat increase, a result that a priori can be explored theoretically.

Some questions are left for future work. In the non-isothermal process, if the final temperature of the bath is not fixed the overdamped approximation is no longer valid, and the final velocity distribution needs to be given by a non-equilibrium distribution coming from an underdamped process. It will be interesting to see what are the modifications.

\section*{Acknowledgement:} This work is supported by the Brazilian agencies CAPES and CNPq. PVP would like to thank FAPERJ for his current PhD fellowship. RA was partially supported by a PhD Fellowship from CAPES. This study was financed in part by Coordena\c c\~ ao de Aperfei\c coamento de Pessoal de N\' ivel Superior - Brasil (CAPES) - Finance Code 001.

\appendix

\section{Calculating the characteristic function}

The definition of the heat distribution is
\begin{equation}
    P(Q) = \langle\delta(Q-Q[x]) \rangle,
\end{equation}
where the average is over the fluctuating degrees of freedom of the system. Here the average is over $v_0,v_\tau,x_0,x_\tau$. Therefore, if the heat depends only on the boundary degrees of freedom we have
\begin{equation}
    P(Q) = \int dv_0dv_\tau dx_0dx_\tau P(v_\tau,v_0,x_\tau,x_0)\delta(Q-Q[x]).
\end{equation}
The joint distribution of the velocities and position can be decomposed as
\begin{equation}
    P(v_\tau,v_0,x_\tau,x_0)=P(v_\tau,v_0)P(x_\tau,x_0),
\end{equation}
since the positions and velocities are independent random variables. And the joint distributions are given by
\begin{equation}
    P(v_\tau,v_0)= P[v_\tau,\tau|v_0,0]P_0(v_0),
\end{equation}
\begin{equation}
    P(x_\tau,x_0)= P[x_\tau,\tau|x_0,0]P_0(x_0),
\end{equation}
where $P_0$ is the initial probability distribution.

In our case, the velocities are assumed to be in an equilibrium stationary distribution at all times, this means that
\begin{equation}
    P[v_\tau,\tau|v_0,0]P_0(v_0) = P_0(v_\tau)P_0(v_0).
\end{equation}
Therefore, the calculation of the heat distribution reduces to
\begin{eqnarray}
    P(Q) = \int dv_0dv_\tau P_0(v_\tau)P_0(v_0)\nonumber\\\int dx_0dx_\tau P[x_\tau,\tau|x_0,0]P_0(x_0) \delta(Q-Q[x]).
\end{eqnarray}
The Dirac delta can be rewritten in terms of its Fourier transform, giving
\begin{eqnarray}
    P(Q) = \int \frac{d\lambda}{2\pi}e^{i\lambda Q} \int dv_0dv_\tau P_0(v_\tau)P_0(v_0)\\\int dx_0dx_\tau P[x_\tau,\tau|x_0,0]P_0(x_0) e^{-i\lambda Q[x]}.
\end{eqnarray}
If $Q[x]$ only depends on the velocity, the integrals in the position reduces to one,
\begin{equation}
    \int dx_0dx_\tau P[x_\tau,\tau|x_0,0]P_0(x_0) = 1,
\end{equation}
letting us with
\begin{eqnarray}
    P(Q) = \int \frac{d\lambda}{2\pi}e^{i\lambda Q}\int dv_0dv_\tau P_0(v_\tau)P_0(v_0)e^{-i\lambda Q[x]},\nonumber\\ 
\end{eqnarray}
one thus can recognize Eq.~\ref{charc}. Where the characteristic function is
\begin{equation}
    Z(\lambda) = \int dv_0dv_\tau P_0(v_\tau)P_0(v_0)e^{-i\lambda Q[x]}.
\end{equation}

\section{Calculation of $P(Q)$}\label{apA}
The Heat distribution can be calculated analytically by the Fourier Transform of the characteristic function.

\begin{equation}
    P(Q) = \int \frac{d\lambda}{2\pi} e^{i\lambda Q} Z(\lambda). 
\end{equation}
The first step is to rewrite the characteristic function eq.~\ref{equili} as
\begin{equation}
    Z(\lambda) = \frac{1}{\sqrt{\alpha(T_f,T_0)}}\frac{1}{\sqrt{1+\left(\lambda+i\frac{\Delta T}{2T_0T_f}\right)^2\alpha(T_f,T_0)^{-1}}}, 
\end{equation}
where $\Delta T = T_f-T_0$ and $\alpha(T_f,T_0) = \left(1+\frac{\Delta T^2}{4T_0T_f}\right)$. By rewriting the characteristic function, the only thing that we need to do is use the formula
\begin{equation}
    \int d\mu \frac{e^{i\mu x}}{\sqrt{1+\mu^2}}  = 2K_0(|x|).
\end{equation}
Thus, by integral transformations one can obtain the probability distribution of eq.~\ref{distheat}.


\begin{thebibliography}{36}%
\makeatletter
\providecommand \@ifxundefined [1]{%
 \@ifx{#1\undefined}
}%
\providecommand \@ifnum [1]{%
 \ifnum #1\expandafter \@firstoftwo
 \else \expandafter \@secondoftwo
 \fi
}%
\providecommand \@ifx [1]{%
 \ifx #1\expandafter \@firstoftwo
 \else \expandafter \@secondoftwo
 \fi
}%
\providecommand \natexlab [1]{#1}%
\providecommand \enquote  [1]{``#1''}%
\providecommand \bibnamefont  [1]{#1}%
\providecommand \bibfnamefont [1]{#1}%
\providecommand \citenamefont [1]{#1}%
\providecommand \href@noop [0]{\@secondoftwo}%
\providecommand \href [0]{\begingroup \@sanitize@url \@href}%
\providecommand \@href[1]{\@@startlink{#1}\@@href}%
\providecommand \@@href[1]{\endgroup#1\@@endlink}%
\providecommand \@sanitize@url [0]{\catcode `\\12\catcode `\$12\catcode
  `\&12\catcode `\#12\catcode `\^12\catcode `\_12\catcode `\%12\relax}%
\providecommand \@@startlink[1]{}%
\providecommand \@@endlink[0]{}%
\providecommand \url  [0]{\begingroup\@sanitize@url \@url }%
\providecommand \@url [1]{\endgroup\@href {#1}{\urlprefix }}%
\providecommand \urlprefix  [0]{URL }%
\providecommand \Eprint [0]{\href }%
\providecommand \doibase [0]{http://dx.doi.org/}%
\providecommand \selectlanguage [0]{\@gobble}%
\providecommand \bibinfo  [0]{\@secondoftwo}%
\providecommand \bibfield  [0]{\@secondoftwo}%
\providecommand \translation [1]{[#1]}%
\providecommand \BibitemOpen [0]{}%
\providecommand \bibitemStop [0]{}%
\providecommand \bibitemNoStop [0]{.\EOS\space}%
\providecommand \EOS [0]{\spacefactor3000\relax}%
\providecommand \BibitemShut  [1]{\csname bibitem#1\endcsname}%
\let\auto@bib@innerbib\@empty
\bibitem [{\citenamefont
  {Barnett}(1946{\natexlab{a}})}]{barnett1946development}%
  \BibitemOpen
  \bibfield  {author} {\bibinfo {author} {\bibfnamefont {M.~K.}\ \bibnamefont
  {Barnett}},\ }\href@noop {} {\bibfield  {journal} {\bibinfo  {journal} {The
  Scientific Monthly}\ }\textbf {\bibinfo {volume} {62}},\ \bibinfo {pages}
  {165} (\bibinfo {year} {1946}{\natexlab{a}})}\BibitemShut {NoStop}%
\bibitem [{\citenamefont
  {Barnett}(1946{\natexlab{b}})}]{barnett1946development2}%
  \BibitemOpen
  \bibfield  {author} {\bibinfo {author} {\bibfnamefont {M.~K.}\ \bibnamefont
  {Barnett}},\ }\href@noop {} {\bibfield  {journal} {\bibinfo  {journal} {The
  Scientific Monthly}\ }\textbf {\bibinfo {volume} {62}},\ \bibinfo {pages}
  {247} (\bibinfo {year} {1946}{\natexlab{b}})}\BibitemShut {NoStop}%
\bibitem [{\citenamefont {Holubec}\ and\ \citenamefont
  {Ryabov}(2021)}]{holubec2021fluctuations}%
  \BibitemOpen
  \bibfield  {author} {\bibinfo {author} {\bibfnamefont {V.}~\bibnamefont
  {Holubec}}\ and\ \bibinfo {author} {\bibfnamefont {A.}~\bibnamefont
  {Ryabov}},\ }\href@noop {} {\bibfield  {journal} {\bibinfo  {journal}
  {Journal of Physics A: Mathematical and Theoretical}\ }\textbf {\bibinfo
  {volume} {55}},\ \bibinfo {pages} {013001} (\bibinfo {year}
  {2021})}\BibitemShut {NoStop}%
\bibitem [{\citenamefont {Ciliberto}(2017)}]{ciliberto2017experiments}%
  \BibitemOpen
  \bibfield  {author} {\bibinfo {author} {\bibfnamefont {S.}~\bibnamefont
  {Ciliberto}},\ }\href@noop {} {\bibfield  {journal} {\bibinfo  {journal}
  {Physical Review X}\ }\textbf {\bibinfo {volume} {7}},\ \bibinfo {pages}
  {021051} (\bibinfo {year} {2017})}\BibitemShut {NoStop}%
\bibitem [{\citenamefont {Mart{\'\i}nez}\ \emph {et~al.}(2016)\citenamefont
  {Mart{\'\i}nez}, \citenamefont {Rold{\'a}n}, \citenamefont {Dinis},
  \citenamefont {Petrov}, \citenamefont {Parrondo},\ and\ \citenamefont
  {Rica}}]{martinez2016brownian}%
  \BibitemOpen
  \bibfield  {author} {\bibinfo {author} {\bibfnamefont {I.~A.}\ \bibnamefont
  {Mart{\'\i}nez}}, \bibinfo {author} {\bibfnamefont {{\'E}.}~\bibnamefont
  {Rold{\'a}n}}, \bibinfo {author} {\bibfnamefont {L.}~\bibnamefont {Dinis}},
  \bibinfo {author} {\bibfnamefont {D.}~\bibnamefont {Petrov}}, \bibinfo
  {author} {\bibfnamefont {J.~M.}\ \bibnamefont {Parrondo}}, \ and\ \bibinfo
  {author} {\bibfnamefont {R.~A.}\ \bibnamefont {Rica}},\ }\href@noop {}
  {\bibfield  {journal} {\bibinfo  {journal} {Nature physics}\ }\textbf
  {\bibinfo {volume} {12}},\ \bibinfo {pages} {67} (\bibinfo {year}
  {2016})}\BibitemShut {NoStop}%
\bibitem [{\citenamefont {Blickle}\ and\ \citenamefont
  {Bechinger}(2012)}]{blickle2012realization}%
  \BibitemOpen
  \bibfield  {author} {\bibinfo {author} {\bibfnamefont {V.}~\bibnamefont
  {Blickle}}\ and\ \bibinfo {author} {\bibfnamefont {C.}~\bibnamefont
  {Bechinger}},\ }\href@noop {} {\bibfield  {journal} {\bibinfo  {journal}
  {Nature Physics}\ }\textbf {\bibinfo {volume} {8}},\ \bibinfo {pages} {143}
  (\bibinfo {year} {2012})}\BibitemShut {NoStop}%
\bibitem [{\citenamefont {Joubaud}\ \emph {et~al.}(2007)\citenamefont
  {Joubaud}, \citenamefont {Garnier},\ and\ \citenamefont
  {Ciliberto}}]{joubaud_fluctuation_2007}%
  \BibitemOpen
  \bibfield  {author} {\bibinfo {author} {\bibfnamefont {S.}~\bibnamefont
  {Joubaud}}, \bibinfo {author} {\bibfnamefont {N.~B.}\ \bibnamefont
  {Garnier}}, \ and\ \bibinfo {author} {\bibfnamefont {S.}~\bibnamefont
  {Ciliberto}},\ }\href {\doibase 10.1088/1742-5468/2007/09/P09018} {\bibfield
  {journal} {\bibinfo  {journal} {J. Stat. Mech.}\ }\textbf {\bibinfo {volume}
  {2007}},\ \bibinfo {pages} {P09018} (\bibinfo {year} {2007})},\ \bibinfo
  {note} {publisher: IOP Publishing}\BibitemShut {NoStop}%
\bibitem [{\citenamefont {Gomez-Solano}\ \emph {et~al.}(2011)\citenamefont
  {Gomez-Solano}, \citenamefont {Petrosyan},\ and\ \citenamefont
  {Ciliberto}}]{gomez-solano_heat_2011}%
  \BibitemOpen
  \bibfield  {author} {\bibinfo {author} {\bibfnamefont {J.~R.}\ \bibnamefont
  {Gomez-Solano}}, \bibinfo {author} {\bibfnamefont {A.}~\bibnamefont
  {Petrosyan}}, \ and\ \bibinfo {author} {\bibfnamefont {S.}~\bibnamefont
  {Ciliberto}},\ }\href {\doibase 10.1103/PhysRevLett.106.200602} {\bibfield
  {journal} {\bibinfo  {journal} {Phys. Rev. Lett.}\ }\textbf {\bibinfo
  {volume} {106}},\ \bibinfo {pages} {200602} (\bibinfo {year} {2011})},\
  \bibinfo {note} {publisher: American Physical Society}\BibitemShut {NoStop}%
\bibitem [{\citenamefont {Imparato}\ \emph {et~al.}(2007)\citenamefont
  {Imparato}, \citenamefont {Peliti}, \citenamefont {Pesce}, \citenamefont
  {Rusciano},\ and\ \citenamefont {Sasso}}]{imparato_work_2007}%
  \BibitemOpen
  \bibfield  {author} {\bibinfo {author} {\bibfnamefont {A.}~\bibnamefont
  {Imparato}}, \bibinfo {author} {\bibfnamefont {L.}~\bibnamefont {Peliti}},
  \bibinfo {author} {\bibfnamefont {G.}~\bibnamefont {Pesce}}, \bibinfo
  {author} {\bibfnamefont {G.}~\bibnamefont {Rusciano}}, \ and\ \bibinfo
  {author} {\bibfnamefont {A.}~\bibnamefont {Sasso}},\ }\href {\doibase
  10.1103/PhysRevE.76.050101} {\bibfield  {journal} {\bibinfo  {journal} {Phys.
  Rev. E}\ }\textbf {\bibinfo {volume} {76}},\ \bibinfo {pages} {050101}
  (\bibinfo {year} {2007})},\ \bibinfo {note} {publisher: American Physical
  Society}\BibitemShut {NoStop}%
\bibitem [{\citenamefont {Imparato}\ \emph {et~al.}(2008)\citenamefont
  {Imparato}, \citenamefont {Jop}, \citenamefont {Petrosyan},\ and\
  \citenamefont {Ciliberto}}]{imparato_probability_2008}%
  \BibitemOpen
  \bibfield  {author} {\bibinfo {author} {\bibfnamefont {A.}~\bibnamefont
  {Imparato}}, \bibinfo {author} {\bibfnamefont {P.}~\bibnamefont {Jop}},
  \bibinfo {author} {\bibfnamefont {A.}~\bibnamefont {Petrosyan}}, \ and\
  \bibinfo {author} {\bibfnamefont {S.}~\bibnamefont {Ciliberto}},\ }\href
  {\doibase 10.1088/1742-5468/2008/10/P10017} {\bibfield  {journal} {\bibinfo
  {journal} {J. Stat. Mech.}\ }\textbf {\bibinfo {volume} {2008}},\ \bibinfo
  {pages} {P10017} (\bibinfo {year} {2008})}\BibitemShut {NoStop}%
\bibitem [{\citenamefont {Chatterjee}\ and\ \citenamefont
  {Cherayil}(2010)}]{chatterjee_exact_2010}%
  \BibitemOpen
  \bibfield  {author} {\bibinfo {author} {\bibfnamefont {D.}~\bibnamefont
  {Chatterjee}}\ and\ \bibinfo {author} {\bibfnamefont {B.~J.}\ \bibnamefont
  {Cherayil}},\ }\href {\doibase 10.1103/PhysRevE.82.051104} {\bibfield
  {journal} {\bibinfo  {journal} {Phys. Rev. E}\ }\textbf {\bibinfo {volume}
  {82}},\ \bibinfo {pages} {051104} (\bibinfo {year} {2010})}\BibitemShut
  {NoStop}%
\bibitem [{\citenamefont {Chatterjee}\ and\ \citenamefont
  {Cherayil}(2011)}]{chatterjee_single-molecule_2011}%
  \BibitemOpen
  \bibfield  {author} {\bibinfo {author} {\bibfnamefont {D.}~\bibnamefont
  {Chatterjee}}\ and\ \bibinfo {author} {\bibfnamefont {B.~J.}\ \bibnamefont
  {Cherayil}},\ }\href {\doibase 10.1088/1742-5468/2011/03/P03010} {\bibfield
  {journal} {\bibinfo  {journal} {J. Stat. Mech.}\ }\textbf {\bibinfo {volume}
  {2011}},\ \bibinfo {pages} {P03010} (\bibinfo {year} {2011})}\BibitemShut
  {NoStop}%
\bibitem [{\citenamefont {Ghosal}\ and\ \citenamefont
  {Cherayil}(2016)}]{ghosal_distribution_2016}%
  \BibitemOpen
  \bibfield  {author} {\bibinfo {author} {\bibfnamefont {A.}~\bibnamefont
  {Ghosal}}\ and\ \bibinfo {author} {\bibfnamefont {B.~J.}\ \bibnamefont
  {Cherayil}},\ }\href {\doibase 10.1088/1742-5468/2016/04/043201} {\bibfield
  {journal} {\bibinfo  {journal} {J. Stat. Mech.}\ }\textbf {\bibinfo {volume}
  {2016}},\ \bibinfo {pages} {043201} (\bibinfo {year} {2016})},\ \bibinfo
  {note} {publisher: IOP Publishing}\BibitemShut {NoStop}%
\bibitem [{\citenamefont {Goswami}(2019)}]{goswami_heat_2019}%
  \BibitemOpen
  \bibfield  {author} {\bibinfo {author} {\bibfnamefont {K.}~\bibnamefont
  {Goswami}},\ }\href {\doibase 10.1103/PhysRevE.99.012112} {\bibfield
  {journal} {\bibinfo  {journal} {Phys. Rev. E}\ }\textbf {\bibinfo {volume}
  {99}},\ \bibinfo {pages} {012112} (\bibinfo {year} {2019})},\ \bibinfo {note}
  {publisher: American Physical Society}\BibitemShut {NoStop}%
\bibitem [{\citenamefont {Paraguass{\'u}}\ and\ \citenamefont
  {Morgado}(2021)}]{paraguassu_heat_2021}%
  \BibitemOpen
  \bibfield  {author} {\bibinfo {author} {\bibfnamefont {P.~V.}\ \bibnamefont
  {Paraguass{\'u}}}\ and\ \bibinfo {author} {\bibfnamefont {W.~A.~M.}\
  \bibnamefont {Morgado}},\ }\href {\doibase 10.1088/1742-5468/abda25}
  {\bibfield  {journal} {\bibinfo  {journal} {J. Stat. Mech.}\ }\textbf
  {\bibinfo {volume} {2021}},\ \bibinfo {pages} {023205} (\bibinfo {year}
  {2021})},\ \bibinfo {note} {publisher: IOP Publishing}\BibitemShut {NoStop}%
\bibitem [{\citenamefont {Chvosta}\ \emph {et~al.}(2020)\citenamefont
  {Chvosta}, \citenamefont {Lips}, \citenamefont {Holubec}, \citenamefont
  {Ryabov},\ and\ \citenamefont {Maass}}]{chvosta_statistics_2020}%
  \BibitemOpen
  \bibfield  {author} {\bibinfo {author} {\bibfnamefont {P.}~\bibnamefont
  {Chvosta}}, \bibinfo {author} {\bibfnamefont {D.}~\bibnamefont {Lips}},
  \bibinfo {author} {\bibfnamefont {V.}~\bibnamefont {Holubec}}, \bibinfo
  {author} {\bibfnamefont {A.}~\bibnamefont {Ryabov}}, \ and\ \bibinfo {author}
  {\bibfnamefont {P.}~\bibnamefont {Maass}},\ }\href {\doibase
  10.1088/1751-8121/ab95c2} {\bibfield  {journal} {\bibinfo  {journal} {J.
  Phys. A: Math. Theor.}\ }\textbf {\bibinfo {volume} {53}},\ \bibinfo {pages}
  {275001} (\bibinfo {year} {2020})}\BibitemShut {NoStop}%
\bibitem [{\citenamefont {Chen}\ \emph {et~al.}(2021)\citenamefont {Chen},
  \citenamefont {Qiu},\ and\ \citenamefont {Quan}}]{chen2021quantum}%
  \BibitemOpen
  \bibfield  {author} {\bibinfo {author} {\bibfnamefont {J.-F.}\ \bibnamefont
  {Chen}}, \bibinfo {author} {\bibfnamefont {T.}~\bibnamefont {Qiu}}, \ and\
  \bibinfo {author} {\bibfnamefont {H.-T.}\ \bibnamefont {Quan}},\ }\href@noop
  {} {\bibfield  {journal} {\bibinfo  {journal} {Entropy}\ }\textbf {\bibinfo
  {volume} {23}},\ \bibinfo {pages} {1602} (\bibinfo {year}
  {2021})}\BibitemShut {NoStop}%
\bibitem [{\citenamefont {Paraguass{\'u}}\ and\ \citenamefont
  {Morgado}(2022)}]{paraguassu2022heat}%
  \BibitemOpen
  \bibfield  {author} {\bibinfo {author} {\bibfnamefont {P.~V.}\ \bibnamefont
  {Paraguass{\'u}}}\ and\ \bibinfo {author} {\bibfnamefont {W.~A.}\
  \bibnamefont {Morgado}},\ }\href@noop {} {\bibfield  {journal} {\bibinfo
  {journal} {Physica A: Statistical Mechanics and its Applications}\ }\textbf
  {\bibinfo {volume} {588}},\ \bibinfo {pages} {126576} (\bibinfo {year}
  {2022})}\BibitemShut {NoStop}%
\bibitem [{\citenamefont {Peliti}\ and\ \citenamefont
  {Pigolotti}(2021)}]{peliti2021stochastic}%
  \BibitemOpen
  \bibfield  {author} {\bibinfo {author} {\bibfnamefont {L.}~\bibnamefont
  {Peliti}}\ and\ \bibinfo {author} {\bibfnamefont {S.}~\bibnamefont
  {Pigolotti}},\ }\href@noop {} {\emph {\bibinfo {title} {Stochastic
  Thermodynamics: An Introduction}}}\ (\bibinfo  {publisher} {Princeton
  University Press},\ \bibinfo {year} {2021})\BibitemShut {NoStop}%
\bibitem [{\citenamefont {Sekimoto}(2010)}]{sekimoto2010stochastic}%
  \BibitemOpen
  \bibfield  {author} {\bibinfo {author} {\bibfnamefont {K.}~\bibnamefont
  {Sekimoto}},\ }\href@noop {} {\emph {\bibinfo {title} {Stochastic
  energetics}}},\ Vol.\ \bibinfo {volume} {799}\ (\bibinfo  {publisher}
  {Springer},\ \bibinfo {year} {2010})\BibitemShut {NoStop}%
\bibitem [{\citenamefont {Seifert}(2012)}]{seifert2012stochastic}%
  \BibitemOpen
  \bibfield  {author} {\bibinfo {author} {\bibfnamefont {U.}~\bibnamefont
  {Seifert}},\ }\href@noop {} {\bibfield  {journal} {\bibinfo  {journal}
  {Reports on progress in physics}\ }\textbf {\bibinfo {volume} {75}},\
  \bibinfo {pages} {126001} (\bibinfo {year} {2012})}\BibitemShut {NoStop}%
\bibitem [{\citenamefont {Nascimento}\ and\ \citenamefont
  {Morgado}(2019)}]{nascimento2019}%
  \BibitemOpen
  \bibfield  {author} {\bibinfo {author} {\bibfnamefont {E.~d.~S.}\
  \bibnamefont {Nascimento}}\ and\ \bibinfo {author} {\bibfnamefont {W.~A.}\
  \bibnamefont {Morgado}},\ }\href@noop {} {\bibfield  {journal} {\bibinfo
  {journal} {EPL (Europhysics Letters)}\ }\textbf {\bibinfo {volume} {126}},\
  \bibinfo {pages} {10002} (\bibinfo {year} {2019})}\BibitemShut {NoStop}%
\bibitem [{\citenamefont {Arold}\ \emph {et~al.}(2018)\citenamefont {Arold},
  \citenamefont {Dechant},\ and\ \citenamefont {Lutz}}]{arold2018heat}%
  \BibitemOpen
  \bibfield  {author} {\bibinfo {author} {\bibfnamefont {D.}~\bibnamefont
  {Arold}}, \bibinfo {author} {\bibfnamefont {A.}~\bibnamefont {Dechant}}, \
  and\ \bibinfo {author} {\bibfnamefont {E.}~\bibnamefont {Lutz}},\ }\href@noop
  {} {\bibfield  {journal} {\bibinfo  {journal} {Physical Review E}\ }\textbf
  {\bibinfo {volume} {97}},\ \bibinfo {pages} {022131} (\bibinfo {year}
  {2018})}\BibitemShut {NoStop}%
\bibitem [{\citenamefont {Paraguass{\'u}}\ \emph {et~al.}(2021)\citenamefont
  {Paraguass{\'u}}, \citenamefont {Aquino},\ and\ \citenamefont
  {Morgado}}]{paraguassu_heat_2021_2}%
  \BibitemOpen
  \bibfield  {author} {\bibinfo {author} {\bibfnamefont {P.~V.}\ \bibnamefont
  {Paraguass{\'u}}}, \bibinfo {author} {\bibfnamefont {R.}~\bibnamefont
  {Aquino}}, \ and\ \bibinfo {author} {\bibfnamefont {W.~A.~M.}\ \bibnamefont
  {Morgado}},\ }\href {http://arxiv.org/abs/2102.09115} {\bibfield  {journal}
  {\bibinfo  {journal} {arXiv:2102.09115 [cond-mat]}\ } (\bibinfo {year}
  {2021})},\ \bibinfo {note} {arXiv: 2102.09115}\BibitemShut {NoStop}%
\bibitem [{\citenamefont {Garc{\'\i}a-Garc{\'\i}a}(2018)}]{garcia2018heat}%
  \BibitemOpen
  \bibfield  {author} {\bibinfo {author} {\bibfnamefont {R.}~\bibnamefont
  {Garc{\'\i}a-Garc{\'\i}a}},\ }\href@noop {} {\bibfield  {journal} {\bibinfo
  {journal} {arXiv preprint arXiv:1812.07311}\ } (\bibinfo {year}
  {2018})}\BibitemShut {NoStop}%
\bibitem [{\citenamefont {Westfall}(2014)}]{westfall2014kurtosis}%
  \BibitemOpen
  \bibfield  {author} {\bibinfo {author} {\bibfnamefont {P.~H.}\ \bibnamefont
  {Westfall}},\ }\href@noop {} {\bibfield  {journal} {\bibinfo  {journal} {The
  American Statistician}\ }\textbf {\bibinfo {volume} {68}},\ \bibinfo {pages}
  {191} (\bibinfo {year} {2014})}\BibitemShut {NoStop}%
\bibitem [{\citenamefont {Leibovich}\ \emph {et~al.}(2016)\citenamefont
  {Leibovich}, \citenamefont {Dechant}, \citenamefont {Lutz},\ and\
  \citenamefont {Barkai}}]{leibovich_aging_2016}%
  \BibitemOpen
  \bibfield  {author} {\bibinfo {author} {\bibfnamefont {N.}~\bibnamefont
  {Leibovich}}, \bibinfo {author} {\bibfnamefont {A.}~\bibnamefont {Dechant}},
  \bibinfo {author} {\bibfnamefont {E.}~\bibnamefont {Lutz}}, \ and\ \bibinfo
  {author} {\bibfnamefont {E.}~\bibnamefont {Barkai}},\ }\href {\doibase
  10.1103/PhysRevE.94.052130} {\bibfield  {journal} {\bibinfo  {journal} {Phys.
  Rev. E}\ }\textbf {\bibinfo {volume} {94}},\ \bibinfo {pages} {052130}
  (\bibinfo {year} {2016})},\ \bibinfo {note} {publisher: American Physical
  Society}\BibitemShut {NoStop}%
\bibitem [{\citenamefont {Hirschberg}\ \emph {et~al.}(2011)\citenamefont
  {Hirschberg}, \citenamefont {Mukamel},\ and\ \citenamefont
  {Sch{\"u}tz}}]{hirschberg_approach_2011}%
  \BibitemOpen
  \bibfield  {author} {\bibinfo {author} {\bibfnamefont {O.}~\bibnamefont
  {Hirschberg}}, \bibinfo {author} {\bibfnamefont {D.}~\bibnamefont {Mukamel}},
  \ and\ \bibinfo {author} {\bibfnamefont {G.~M.}\ \bibnamefont {Sch{\"u}tz}},\
  }\href {\doibase 10.1103/PhysRevE.84.041111} {\bibfield  {journal} {\bibinfo
  {journal} {Phys. Rev. E}\ }\textbf {\bibinfo {volume} {84}},\ \bibinfo
  {pages} {041111} (\bibinfo {year} {2011})},\ \bibinfo {note} {publisher:
  American Physical Society}\BibitemShut {NoStop}%
\bibitem [{\citenamefont {Aghion}\ \emph {et~al.}(2019)\citenamefont {Aghion},
  \citenamefont {Kessler},\ and\ \citenamefont
  {Barkai}}]{aghion_non-normalizable_2019}%
  \BibitemOpen
  \bibfield  {author} {\bibinfo {author} {\bibfnamefont {E.}~\bibnamefont
  {Aghion}}, \bibinfo {author} {\bibfnamefont {D.~A.}\ \bibnamefont {Kessler}},
  \ and\ \bibinfo {author} {\bibfnamefont {E.}~\bibnamefont {Barkai}},\ }\href
  {\doibase 10.1103/PhysRevLett.122.010601} {\bibfield  {journal} {\bibinfo
  {journal} {Phys. Rev. Lett.}\ }\textbf {\bibinfo {volume} {122}},\ \bibinfo
  {pages} {010601} (\bibinfo {year} {2019})}\BibitemShut {NoStop}%
\bibitem [{\citenamefont {Dechant}\ \emph {et~al.}(2012)\citenamefont
  {Dechant}, \citenamefont {Lutz}, \citenamefont {Kessler},\ and\ \citenamefont
  {Barkai}}]{dechant_superaging_2012}%
  \BibitemOpen
  \bibfield  {author} {\bibinfo {author} {\bibfnamefont {A.}~\bibnamefont
  {Dechant}}, \bibinfo {author} {\bibfnamefont {E.}~\bibnamefont {Lutz}},
  \bibinfo {author} {\bibfnamefont {D.~A.}\ \bibnamefont {Kessler}}, \ and\
  \bibinfo {author} {\bibfnamefont {E.}~\bibnamefont {Barkai}},\ }\href
  {\doibase 10.1103/PhysRevE.85.051124} {\bibfield  {journal} {\bibinfo
  {journal} {Phys. Rev. E}\ }\textbf {\bibinfo {volume} {85}},\ \bibinfo
  {pages} {051124} (\bibinfo {year} {2012})},\ \bibinfo {note} {publisher:
  American Physical Society}\BibitemShut {NoStop}%
\bibitem [{\citenamefont {Barkai}\ \emph {et~al.}(2014)\citenamefont {Barkai},
  \citenamefont {Aghion},\ and\ \citenamefont {Kessler}}]{barkai_area_2014}%
  \BibitemOpen
  \bibfield  {author} {\bibinfo {author} {\bibfnamefont {E.}~\bibnamefont
  {Barkai}}, \bibinfo {author} {\bibfnamefont {E.}~\bibnamefont {Aghion}}, \
  and\ \bibinfo {author} {\bibfnamefont {D.~A.}\ \bibnamefont {Kessler}},\
  }\href {\doibase 10.1103/PhysRevX.4.021036} {\bibfield  {journal} {\bibinfo
  {journal} {Phys. Rev. X}\ }\textbf {\bibinfo {volume} {4}},\ \bibinfo {pages}
  {021036} (\bibinfo {year} {2014})}\BibitemShut {NoStop}%
\bibitem [{\citenamefont {Kessler}\ and\ \citenamefont
  {Barkai}(2010)}]{kessler_infinite_2010}%
  \BibitemOpen
  \bibfield  {author} {\bibinfo {author} {\bibfnamefont {D.~A.}\ \bibnamefont
  {Kessler}}\ and\ \bibinfo {author} {\bibfnamefont {E.}~\bibnamefont
  {Barkai}},\ }\href {\doibase 10.1103/PhysRevLett.105.120602} {\bibfield
  {journal} {\bibinfo  {journal} {Phys. Rev. Lett.}\ }\textbf {\bibinfo
  {volume} {105}},\ \bibinfo {pages} {120602} (\bibinfo {year}
  {2010})}\BibitemShut {NoStop}%
\bibitem [{\citenamefont {Ray}\ and\ \citenamefont
  {Reuveni}(2020)}]{ray_diffusion_2020}%
  \BibitemOpen
  \bibfield  {author} {\bibinfo {author} {\bibfnamefont {S.}~\bibnamefont
  {Ray}}\ and\ \bibinfo {author} {\bibfnamefont {S.}~\bibnamefont {Reuveni}},\
  }\href {\doibase 10.1063/5.0010549} {\bibfield  {journal} {\bibinfo
  {journal} {J. Chem. Phys.}\ }\textbf {\bibinfo {volume} {152}},\ \bibinfo
  {pages} {234110} (\bibinfo {year} {2020})},\ \bibinfo {note} {publisher:
  American Institute of Physics}\BibitemShut {NoStop}%
\bibitem [{\citenamefont {Ryabov}(2015)}]{ryabov2015stochastic}%
  \BibitemOpen
  \bibfield  {author} {\bibinfo {author} {\bibfnamefont {A.}~\bibnamefont
  {Ryabov}},\ }\href@noop {} {\emph {\bibinfo {title} {Stochastic dynamics and
  energetics of biomolecular systems}}}\ (\bibinfo  {publisher} {Springer},\
  \bibinfo {year} {2015})\BibitemShut {NoStop}%
\bibitem [{\citenamefont {Abramowitz}\ \emph {et~al.}(1988)\citenamefont
  {Abramowitz}, \citenamefont {Stegun},\ and\ \citenamefont
  {Romer}}]{abramowitz1988handbook}%
  \BibitemOpen
  \bibfield  {author} {\bibinfo {author} {\bibfnamefont {M.}~\bibnamefont
  {Abramowitz}}, \bibinfo {author} {\bibfnamefont {I.~A.}\ \bibnamefont
  {Stegun}}, \ and\ \bibinfo {author} {\bibfnamefont {R.~H.}\ \bibnamefont
  {Romer}},\ }\href@noop {} {\enquote {\bibinfo {title} {Handbook of
  mathematical functions with formulas, graphs, and mathematical tables},}\ }
  (\bibinfo {year} {1988})\BibitemShut {NoStop}%
\bibitem [{\citenamefont {Jarzynski}\ and\ \citenamefont
  {W{\'o}jcik}(2004)}]{jarzynski2004classical}%
  \BibitemOpen
  \bibfield  {author} {\bibinfo {author} {\bibfnamefont {C.}~\bibnamefont
  {Jarzynski}}\ and\ \bibinfo {author} {\bibfnamefont {D.~K.}\ \bibnamefont
  {W{\'o}jcik}},\ }\href@noop {} {\bibfield  {journal} {\bibinfo  {journal}
  {PRL}\ }\textbf {\bibinfo {volume} {92}},\ \bibinfo {pages} {230602}
  (\bibinfo {year} {2004})}\BibitemShut {NoStop}%
\end{thebibliography}
\end{document}